\documentclass[12pt]{article}
\usepackage{amsfonts}
\usepackage{amsmath}
\usepackage{amssymb}
\usepackage{graphicx}
\usepackage{float}
\usepackage{color}
\textwidth=6.5in \hoffset=-.75in \voffset=-1in \numberwithin{equation}{section}
\numberwithin{figure}{section}

\newcommand {\be}{\begin{equation}}
\newcommand {\ee}{\end{equation}}
\begin{document}

\begin{titlepage}
\vspace{1cm}
\begin{center}
{\Large \bf {Exact helicoidal and catenoidal solutions in Einstein-Maxwell theory}}\\
\end{center}
\vspace{2cm}
\begin{center}
{A.M. Ghezelbash \footnote{ E-Mail: amg142@campus.usask.ca}, Vineet Kumar \footnote{E-Mail: v.kumar@usask.ca}}
\\
Department of Physics and Engineering Physics, \\ University of Saskatchewan, \\
Saskatoon, Saskatchewan S7N 5E2, Canada\\
\vspace{1cm}
\vspace{2cm}
\end{center}

\begin{abstract}
We present several new exact solutions in five and higher dimensional Einstein-Maxwell theory by embedding the Nutku instanton.  The metric functions for the five-dimensional solutions depend only on {a} radial coordinate and on two spatial coordinates for the {six} and higher dimensional solutions. The six and higher dimensional metric functions are convoluted-like integrals of two special functions. We find that the solutions are {regular almost} everywhere and some spatial sections of the solution {describe wormhole} handles. We also find a class of exact and nonstationary convoluted-like solutions to the Einstein-Maxwell theory with a cosmological constant.

\end{abstract}
\end{titlepage}\onecolumn 
\bigskip 

\section{Introduction}
Gravitational instantons are the regular and complete Euclidean signature solutions, with self-dual curvature two-form, to the Einstein field equations in vacuum \cite{grin} or with a cosmological constant term \cite{Gib78}.  There are several well known solutions such as Taub-NUT \cite{grinTN}, Eguchi-Hanson \cite{grinEH} and Atiyah-Hitchin metrics \cite{grinAH}.  The instanton solutions, in general, are the result of reduction the complex elliptic Monge-Amp\`ere  equation \cite{Mong} on a complex manifold of dimension 2 to only one real variable. 
These solutions play an important role  in construction of higher-dimensional solutions to extended theories of gravity \cite{biglist1} and supergravity \cite{biglist3}, \cite{biglist31} as well as quantum properties of the black holes \cite{biglist32}.  In fact, these self-dual geometries have been used in \cite{che} to construct the M-theory realizations of the fully localized D2 branes in type IIA string theory that {intersect the D6 branes}. The M-theory solutions involve the convoluted-like integrals of two special functions and upon compactification over a compact coordinate of the transverse self-dual geometry, yield the supersymmetric solutions for the fully localized intersecting branes. One main feature of the solutions is that they are valid near and far from the core of D6 branes. Moreover, inspired by the convoluted-like structure of membrane solutions, some new convoluted solutions to the six and higher dimensional Einstein-Maxwell theory were constructed and {studied} in \cite{Gh}.  The convoluted solutions even can be generalized to include the cosmological constant in six and higher-dimensional Einstein-Maxwell theory.

The gravitational instantons also are the dominant metrics in path-integral formulation of Euclidean quantum gravity and {are closely} related to {minimal} surfaces in Eulidean space \cite{Nut96}.  In fact, the equations for any {2-dimensional} minimal surfaces, provide a solution to the real elliptic  Monge-Amp\`ere  equation on a real manifold of {2 dimensions}. These solutions then lead to the K\"ahler metrics for some gravitational instantons beside the {aforementioned well-known} solutions.

Almost all research in gravitational instantons and Einstein-Maxwell theory has been focused exclusively on spherically symmetric solutions so far. In this article, we draw attention to gravitational instantons of novel geometries constructed from minimal surfaces by Nutku \cite{Nut96} and embed them to generate a new class of exact helicoidal and catenoidal solutions in Einstein-Maxwell Theory.
Inspired by the existence of such gravitational instantons in four dimensions and the well-known methods to construct the higher-dimensional convoluted solutions to the Einstein-Maxwell theory, in this article, we construct and study new class of exact helicoidal and catenoidal solutions in Einstein-Maxwell theory.  We note that to our knowledge, these solutions which are generated from helicoid and catenoid minimal surfaces, have not been studied before in Einstein-Maxwell Theory.

The article is organized as follows. 
We begin with a brief overview of minimal surfaces and instantons  and present the metric for the Nutku gravitational instantons in section 2.  In section 3, we embed the Nutku metrics corresponding to helicoid and catenoid cases to get exact solutions in 5 dimensions. In section 4, we construct convolution-like general solutions for the Einstein-Maxwell theory in six dimensions and discuss the solutions. In section 5, we generalize the convoluted-like solutionst for the Einstein-Maxwell theory in any dimensions greater than six and discuss the solutions. Finally in section 6,  by using a very special separation of variables in the metric function, we find the most general cosmological solutions to the Einstein-Maxwell theory in presence of a cosmological constant. We wrap up the article by concluding remarks.

\label{sec:Intro}

\section{Minimal surfaces and the four dimensional Nutku instantons}
\label{sec:Nutku4D}
The study of minimal surfaces began with Lagrange's question on the existence of surfaces that minimize the area, subject to some boundary constraints. Physically, they represent soap films on wire frames. There are several equivalent mathematical definitions of minimal surfaces. We state two important ones \cite{Mee12}:
1) A surface $M\subset \mathbb{R}^3$ is minimal if and only if its mean curvature vanishes identically.  2) A surface $M\subset \mathbb{R}^3$ is minimal if and only if it is a critical point of the area functional for all compactly supported versions.
The plane, the helicoid, the catenoid, the gyroid, Scherk surface, Enneper surface and Costa's surface are some examples \cite{Fom05}. In $\mathbb{R}^3$, the helicoid and the plane are the only ruled 2-surfaces \cite{Mee12}, i.e. surfaces generated by the rotation of a line. The helicoid and the catenoid are locally isometric \cite{seven} and are {conjugates} of each other \cite{Mee12}.\footnote{{Conjugate surfaces have the interesting property that a straight line in one surface can be mapped to a geodesic in the other and vice-versa.}}

Observing that nearly all well-known solutions in Einstein-Maxwell theory have spherical symmetry, this article highlights the role of  minimal surfaces and instantons in constructing new solutions of novel geometries. Instantons are pseudo-particles which first appeared in Yang-Mills theory as minimum-action, classical solutions in Euclidean\footnote{Curiously, Wick rotation ($t \rightarrow i t$) plays an important role in instanton physics in both quantum theory and general relativity.} spacetime \cite{Bel75}. Their discovery inspired the notion of gravitational instantons, which soon found use in Schwarzschild black hole radiance calculations through Euclideanization \cite{Har76}.
In 1978, Comtet \cite{Com78} showed that the multi-BPST-instanton solution of Witten \cite{Wit77} corresponds to minimal surfaces.  In a similar vein, Nutku \cite{Nut96} proved that for every minimal surface in $\mathbb{R}^3$, there is a gravitational instanton with anti-self-dual curvature and gave some explicit metrics which we will embed in higher dimensions.

A class of gravitational instantons may be representated by the metric
\begin{equation}
ds^2=\frac{1}{P(t,x)}\{(f_t^2+\kappa)(dt^2+dy^2)+(1+f_x^2)(dx^2+dz^2)+2f_tf_x(dtdx+dydz)\}, 
\end{equation}
where $P(t,x)=\sqrt{1+\kappa f_t^2+f_x^2}$, $f_t=\frac{\partial f(t,x)}{\partial t}$ and $f_x=\frac{\partial f(t,x)}{\partial x}$, if the function $f(t,x)$ satisfies the quasi-linear, elliptic-hyperbolic partial differential equation 
\begin{equation}
(1+f_x^2)f_{tt}-2f_tf_xf_{tx}+(\kappa+f_t^2)f_{xx}=0\label{feq}.
\end{equation}

The ``Lagrange equation" (\ref{feq}) with $\kappa=1$ defines minimal surfaces in $\mathbb{R}^3$. For $\kappa=-1$, it reduces to the Born-Infeld equation \cite{Nut96}, which arises in a non-linear generalization of Maxwell electrodynamics. These two cases are also related through a Wick rotation ($t\rightarrow{{i}}t$) \cite{Dey03}. Interestingly, the Born-Infeld equation is also related to the maximal surface equation in Lorentz-Minkowski space $\mathbb{L}^3$ by a Wick rotation \cite{Dey17}. Since the general metric above provides a large class of solutions, we will restrict our attention to the helicoid-catenoid solutions. Noting that the helicoid represented by $f(t,x)=a\, {tan}^{-1}(\frac{x}{t})$ is a solution to equation (\ref{feq}), after substitution and subsequent coordinate transformations $x=r \cos {\theta}$ and $t= r \sin {\theta}$, we obtain the Nutku helicoid instanton as
\be
ds^2=\frac{{d{{r}}}^{2}+ \left( {a}^{2}+{r}^{2} \right) {d{{\theta}}}^{2}+
 \left( 1+{\frac {{a}^{2} \left( \sin \left( \theta \right)  \right) ^
{2}}{{r}^{2}}} \right) {d{{y}}}^{2}-{\frac {{a}^{2}\sin \left( 2\,
\theta \right) d{{y}}d{{z}}}{{r}^{2}}}+ \left( 1+{\frac {{a}^{2}
 \left( \cos \left( \theta \right)  \right) ^{2}}{{r}^{2}}} \right) {d
{{z}}}^{2}
}{ \sqrt{1+{\frac {{a}^{2}}{{r}^{2}}}}}\label{hel},
\ee
where we consider {$0<r<\infty$, $0\leq\theta\leq2\pi$ and $y$ and $z$ could be considered the periodic coordinates on the 2-torus} \cite{Ali99}.  In \cite{ADDED}, the authors studied the solutions of the Dirac equation in the background of the metric (\ref{hel}) and its singularities.
The metric (\ref{hel}) is asymptotically Euclidean and would correspond to a catenoidal solution if $a^2$ is replaced with $-a^2$. From here on, we will use $\epsilon=\pm 1$ to differentiate between the helicoid ($\epsilon=1$) and catenoid ($\epsilon=- 1$) cases, respectively.
\be
ds_{Nutku}^2=\frac{{d{{r}}}^{2}+ \left( \epsilon{a}^{2}+{r}^{2} \right) {d{{\theta}}}^{2}+
	\left( 1+{\epsilon\frac {{a}^{2} \left( \sin \left( \theta \right)  \right) ^
			{2}}{{r}^{2}}} \right) {d{{y}}}^{2}-{\epsilon\frac {{a}^{2}\sin \left( 2\,
			\theta \right) d{{y}}d{{z}}}{{r}^{2}}}+ \left( 1+{\epsilon\frac {{a}^{2}
			\left( \cos \left( \theta \right)  \right) ^{2}}{{r}^{2}}} \right) {d
		{{z}}}^{2}
}{ \sqrt{1+{\epsilon\frac {{a}^{2}}{{r}^{2}}}}}\label{Nutku}.
\ee
 The Kretchmann invariant of the helicoid or catenoid instanton is given by 
\begin{equation}
{\cal K}=\frac{72a^4}{r^4(r^2+\epsilon a^2)^2}+\frac{24a^8}{r^6(r^2+\epsilon a^2)^3}.
\end{equation}
The helicoid has only a curvature singularity at $r=0$, while for the catenoid, there is another singularity at $r=a$.

\section{Helicoid and catenoid solutions in 5-dimensional Einstein-Maxwell theory}
\label{sec:Nutku5D}

We consider the source-free Einstein-Maxwell equations with geometrized units in $D$-dimensions that are given by
\begin{eqnarray}
R_{\mu\nu}&=&F_{\mu}^{\lambda}F_{\nu\lambda}-\frac{1}{4+2(D-4)}g_{\mu\nu}F^2\label{eq1},\\
F^{\mu\nu}_{;\mu}&=&0\label{eq2},
\end{eqnarray}
where the electromagnetic field tensor is given by
\be
F_{\mu\nu}=A_{\mu,\nu}-A_{\nu,\mu},\label{eq3}
\ee
in terms of the electromagnetic potential $A^{\mu}$. {We will take its only non-zero component as}
\be
{A_t}={\sqrt{\frac{D-2}{D-3}}\frac{1}{H(r)}}\label{gaugeD}.
\ee 

In {5 dimensions}, we consider the following  $5D$ ansatz by adding a time coordinate to the Nutku helicoid or catenoid instanton
\be
\begin{split}
	ds^2 &=-{\frac {d{t}^{2}}{ \left( H \left( r \right)  \right) ^{2}}}\\
	&+{\frac{H(r)}{ \sqrt{1+{\frac 
					{{a}^{2}}{{r}^{2}}}}}  \left( {d_{{r}}}^{2}+ \left( {a}^{2}+{r}^{2}
		\right) {d_{{\theta}}}^{2}
		+\left( 1+{\frac {{a}^{2} \left( \sin
				\left( \theta \right)  \right) ^{2}}{{r}^{2}}} \right) {d_{{y}}}^{2}-
		{\frac {{a}^{2}\sin \left( 2\,\theta \right) d_{{y}}d_{{z}}}{{r}^{2}}}
		+ \left( 1+{\frac {{a}^{2} \left( \cos \left( \theta \right)  \right) 
				^{2}}{{r}^{2}}} \right) {d_{{z}}}^{2} \right)}.
\end{split}\label{metr5}
\ee

We find that all Einstein and Maxwell equations are satisfied if the metric function $H(r)$ satisfies the differential equation
\be
 \epsilon \left( {\frac {{\rm d}^{2}}{{\rm d}{r}^{2}}}H \left( r \right) 
 \right) {a}^{2}+ \left( {\frac {{\rm d}^{2}}{{\rm d}{r}^{2}}}H
 \left( r \right)  \right) {r}^{2}+ \left( {\frac {\rm d}{{\rm d}r}}H
 \left( r \right)  \right) r
=0.\label{ode}
\ee
The solutions to (\ref{ode}) are given by
\be
H(r)=
{c_1}+{C_2}\ln(\frac{r+\sqrt{\epsilon a^2+r^2}}{a}),\label{sol5}
\ee
where $c_1$ and $C_2$ are two constants. 

The logarithmic solutions above correspond to {$\sinh^{-1}(\frac{r}{a})$} and {$\cosh^{-1}(\frac{r}{a})$} functions. Interestingly, these functions appear in the study of collapsing catenoidal soap films \cite{soap} and in the embedding of wormhole handles \cite{Har93}.  We note that a submanifold of the $5D$ helicoidal spacetime is conformally equivalent to a wormhole handle (${d{{r}}}^{2}+ \left( {a}^{2}+{r}^{2} \right) {d{{\theta}}}^{2}$) \cite{Har93}\cite{Ell73}.

\begin{figure}[H]
\centering
\includegraphics[width=0.4\textwidth]{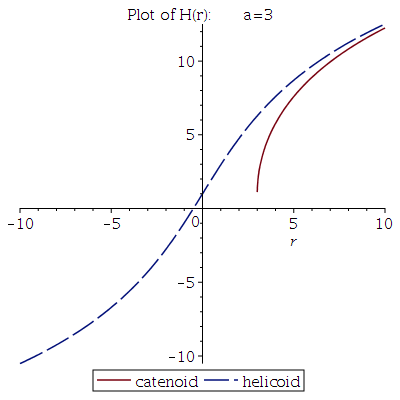}
\caption{$H(r)$, for the helicoid and catenoid solutions, with $a=3,c_1=1,c_2=6$.}
\label{fig1}
\end{figure}

Inspired by the soap film solutions  \cite{soap}, we set $c_1=1$ and {$C_2=c_2\, a$},\footnote{{In fact, we may even get rid of the constant parameter $c_2$ by setting it to be $c_2=2$ in order to match the height function for a catenoidal soap film. It is useful to keep $c_2$ for now to illustrate what it may do to the electric field.}} so that in the limit $a\rightarrow 0$, the metric (\ref{metr5}) becomes Minkowski spacetime. {Thus, we have}\\ 
\begin{align}
H(r)=1+c_2\, a \sinh^{-1}(\frac{r}{a}),\\
H(r)=1+c_2\, {a} \cosh^{-1}(\frac{r}{a}),
\end{align}
{for the helicoid and the catenoid solutions, respectively. Figure \ref{fig1} shows the typical behaviour of the metric functions $H(r)$ for helicoid and catenoid solutions.} It is important to note here that in case of the helicoid, the function $H(r)$ can become negative and hence change the metric signature from Lorentzian to Riemannian at some radial coordinate $r$.  Although such signature changes are of interest in quantum cosmology \cite{Ell92}, and may be dealt with through a Wick rotation, we wish to keep the metric well-behaved and thus require {$a\geq 0$ and $c_2\geq 0$}.

\begin{figure}[H]
\centering
\includegraphics[width=0.4\textwidth]{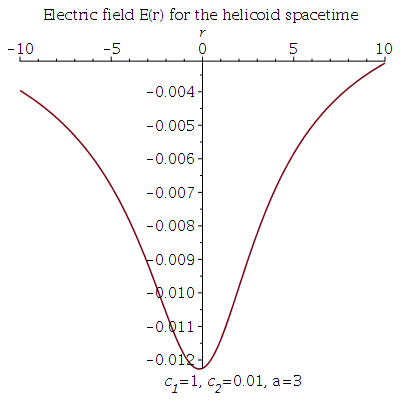}
\includegraphics[width=0.4\textwidth]{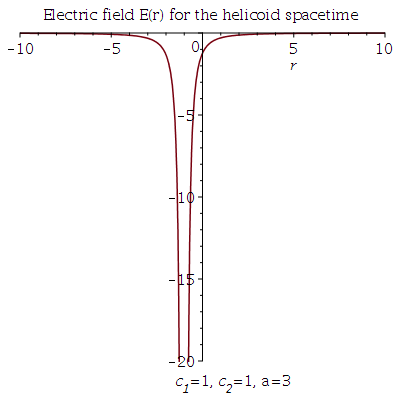}
\caption{The $r$-component of the electric field as a function of $r$, for the helicoid solution with $a=3,c_1=1,c_2=0.01$ (left) and $a=2,c_1=1,c_2=1$ (right).}
\label{fig2}
\end{figure}

Figure \ref{fig2} shows how the electric field may or may not diverge in the helicoid spacetime, depending on the choice of constants. The region $r<0$ is included in the {plots only} to illustrate this behaviour.

\begin{figure}[H]
\centering
\includegraphics[width=0.4\textwidth]{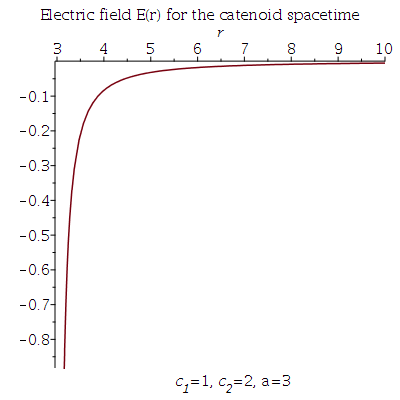}
\caption{$E_r(r)$ for the catenoid solution with $a=3,c_1=1,c_2=2$.}
\label{E_cat}
\end{figure}

The electric field for the catenoid spacetime ($a<r<\infty$) is shown in Figure \ref{E_cat}.

\begin{figure}[H]
	\centering
	\includegraphics[width=0.4\textwidth]{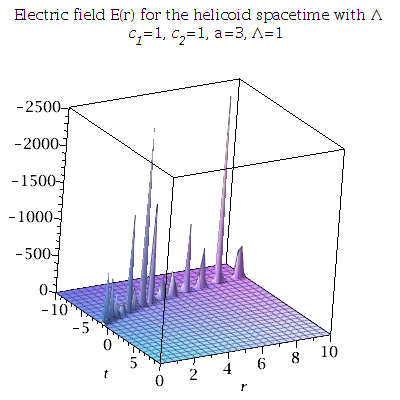}
	\includegraphics[width=0.4\textwidth]{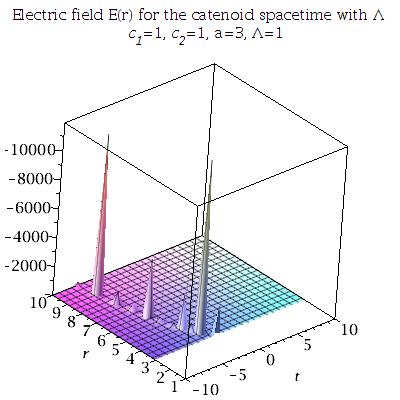}
	\caption{$E_r(r,t)$ with $a=3,c_1=1,c_2=1, \Lambda=1$, for the helicoid solution (left) and at the catenoid solution (right).}
	\label{cosmo5D}
\end{figure}

If we replace $H(r)$ with $H(r,t)$ in the metric ansatz (\ref{metr5}), we can get a dynamic solution to the Einstein-Maxwell equations with a cosmological constant $\Lambda$, which is given by
\be
H(r)={1}\pm\sqrt{\Lambda}t+{c_2}{\,a}\ln(\frac{r+\sqrt{\epsilon a^2+r^2}}{a}).
\ee
Again, if we are to avoid the {metric} signature change while keeping {$a$} and $c_2$ positive, we should only consider the $+\sqrt{\Lambda}t$ solution. Figure \ref{cosmo5D} illustrates {the electric field as a function of r and t for the helicoid and catenoid spacetimes with a cosmological constant.}

\section{The convoluted solutions in six dimensional Einstein-Maxwel theory based on embedded Nutku instantons}
\label{sec:Nutku6D}
\begin{figure}[H]
\centering
\includegraphics[width=0.4\textwidth]{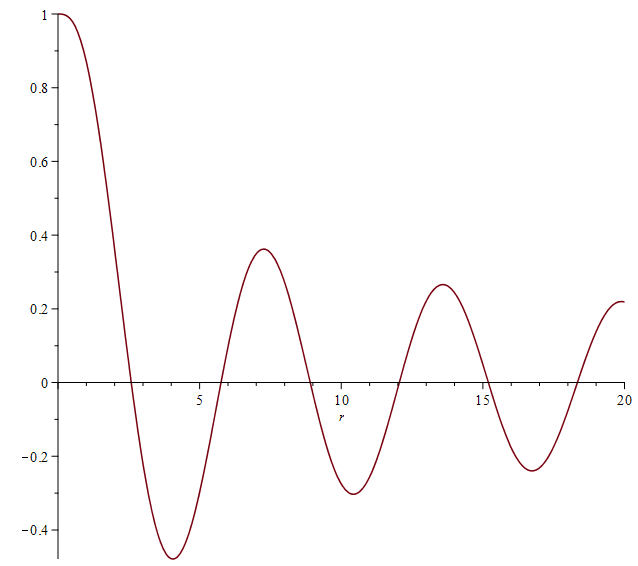}
\includegraphics[width=0.4\textwidth]{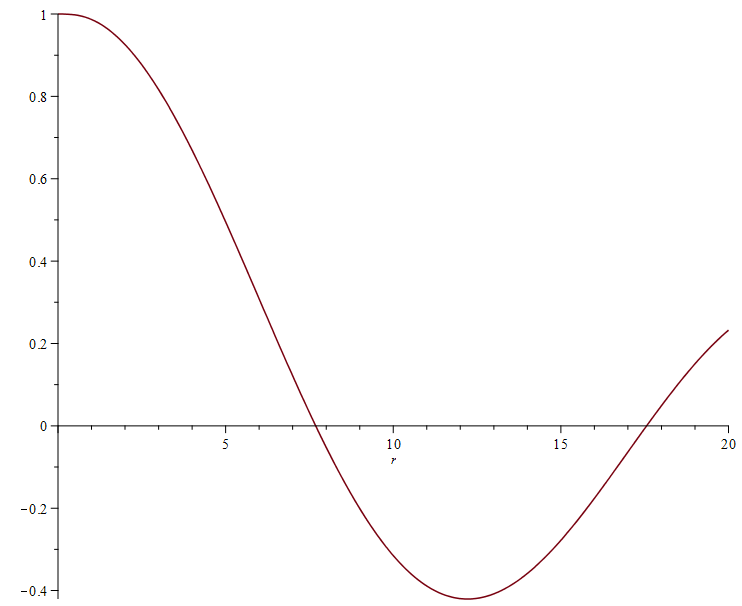}
\caption{The function ${\cal H}_D\left( 0,0,{a}^{2}c^2,0,\sqrt{{\frac {{a}^{2}+{r}^{2}}{{r}^{2}}}}\right)$ versus $r$, where  we set $a=1,c^2=1$ (left) and $a=1,c^2=0.1$ (right).}
\label{fig3}
\end{figure}

We consider the six-dimensional metric ansatz 
\be
ds^{2}=-H(r,x)^{-2}dt^{2}+H(r,x)^{2/3}(dx^2+ds_{Nutku}^2),\label{N6D}
\ee
where $ds_{Nutku}^2$ is the four-dimensional Nutku space with parameter $a$ which is given by (\ref{Nutku}). We also consider the Maxwell gauge field as
\be
A_t(r,x)=\frac{2\sqrt{3}}{3H(r,x)}.\label{g}
\ee
The gravitational Einstein's field equations and the Maxwell equations provide that the metric function $H(r,x)$ must satisfy the partial differential equation
\be
{r}^{2} \sqrt{{\frac {{a}^{2}+{r}^{2}}{{r}^{2}}}}{\frac {\partial ^{2}
}{\partial {x}^{2}}}H \left( r,x \right) + (r^2+ {a}^{2})\left( {\frac {\partial ^{2
}}{\partial {r}^{2}}}H \left( r,x \right)  \right)
+ \left( {\frac {\partial }{\partial r}}H \left( r,x \right) 
 \right) r=0.\label{PD}
\ee
We can solve the partial differential equation (\ref{PD}) by separating the variables as
$H(r,x)=1+R(r)X(x)$. We find two ordinary differential equations for $R(x)$ and $X(x)$ that are given by 
\be
{\frac {{\rm d}^{2}}{{\rm d}{r}^{2}}}R
 \left( r \right) +\epsilon{c^2 {R} \left( r \right)  \left(  \sqrt{1+{\frac {{a}^{
2}}{{r}^{2}}}} \right) ^{-1}}+{\frac { \left( {\frac {\rm d}{{\rm d}r}
}R \left( r \right)  \right) r}{{a}^{2}+{r}^{2}}} =0,
\label{eqR}\ee
and
\be
{\frac {{ d}^{2}}{{d}{x}^{2}}}X \left( x \right) -\epsilon c^2 {X}
 \left( x \right) =0,\label{eqX}
\ee
respectively. We first consider $\epsilon=+1$, where the 
the solutions to (\ref{eqR}) are given by
\be
\begin{split}
R(r) &=A_{{1}}{ {\cal H}_D} \left( 0,0,{a}^{2}c^2,0,\sqrt {{\frac {{a}^{2}+{r}^{2
}}{{r}^{2}}}} \right) \\
&+A_{{2}}{{\cal H}_D} \left( 0,0,{a}^{2}c^2,0,\sqrt 
{{\frac {{a}^{2}+{r}^{2}}{{r}^{2}}}} \right) \int ^{\sqrt {{\dfrac {{a}
^{2}+{r}^{2}}{{r}^{2}}}}}\!{\frac {1}{ \left( {{f}}^{2}-1
 \right)  \left( {{\cal H}_D} \left( 0,0,{a}^{2}c^2,0,{f} \right) 
 \right) ^{2}}}{df},
\end{split}\label{Rsol}
\ee
in terms of Heun-$D$ functions ${{\cal H}_D} $ and $A_1$ and $A_2$ are two constants.  We note that the Heun-$D$ functions  ${\cal H}_D\left( \alpha,\beta,\gamma,\delta,z \right) $  are the solutions to the Heun double confluent equation
\be
{\frac {{\rm d}^{2}}{{\rm d}{z}^{2}}}y(z) -{\frac {
 \left( \alpha\,{z}^{4}-2\,{z}^{5}+4\,{z}^{3}-\alpha-2\,z \right) }{ \left( z-1 \right) ^{3}
 \left( z+1 \right) ^{3}}}{
\frac {\rm d}{{\rm d}z}}y(z)={\frac { \left( -{z}^{2}\beta+ \left( -
\gamma-2\,\alpha \right) z-\delta \right) \ }{
 \left( z-1 \right) ^{3} \left( z+1 \right) ^{3}}}y(z),
\ee
with the boundary conditions $y(0)=1$, $ {\dfrac {d}{{d}z}}y(0)=0$. In figure \ref{fig3}, we show the typical behaviour of the Heun-$D$ function for a few different values of the separation constant $c$.

The solutions to (\ref{eqX}) with $\epsilon=+1$ for $X(x)$ are given by
\be
X(x)=B_{1}\,{{\rm e}^{ {c}x}}+B_{2}\,{{\rm e}^{-{c}x}}\label{Xsol},
\ee
where $B_1$ and $B_2$ are two constants. We can superimpose all the solutions (\ref{Rsol}) (where we choose $A_2=0$) and (\ref{Xsol}) with different values for the separation constant $c$, to construct the most general convoluted solutions to the partial differential equation (\ref{PD}), in the form
\be
H(r,x)=1+\int_{0}^{\infty }{\cal H}_D\left(0,0,{a}^{2}c^2,0,\sqrt{{\frac {{a}^{2}+{r}^{2}}{{r}^{2}}}}\right)(f(c){{e}^{cx}}+g(c){{e}^{-cx}})\,{d}c,\label{Hc}
\ee
where $f(c)$ and $g(c)$ are two arbitrary functions in terms of separation constant $c$. To fix the arbitrary functions $f(c)$ and $g(c)$, we note that in the limit of $a\rightarrow 0$, the Nutku space describes a four-dimensional space $D^2 \otimes T^2$ with the line element
\be
ds_{a=0}^2=dr^2+r^2d\theta^2+dy^2+dz^2.\label{Na0}
\ee
Quite interestingly, in this limit, we find an exact analytical {solution} to six-dimensional Einstein-Maxwell theory with the line element  
\be
d{\cal S}^{2}=-H_0(r,x)^{-2}dt^{2}+H_0(r,x)^{2/3}(dx^2+ds_{a=0}^2),\label{N6Da0}
\ee
and the Maxwell gauge field 
\be
{\cal A}_t(r,x)=\frac{2\sqrt{3}}{3H_0(r,x)},\label{ga0}
\ee
where the exact analytic metric function $H_0(r,x)$ is
\be
H_0(r,x)=1+\frac{h_0}{\sqrt{r^2+x^2}}\label{H0},
\ee
and $h_0$ is a constant. We can now fix the functions $f(c)$ and $g(c)$ by requiring that the metric (\ref{N6D}) and the gauge field (\ref{g}), must approach to the exact analytical metric (\ref{N6Da0}) and the gauge field (\ref{ga0}), respectively, in the limit of $a \rightarrow 0$. These requirements imply that the 
convoluted metric function $H(r,x)$ in equation (\ref{Hc}), must be equal to the exact analytic metric function $H_0(r,x)$ in equation (\ref{H0}), in the limit of $a \rightarrow 0$. The  integrand of the convoluted metric function $H(r,x)$ in equation (\ref{Hc}) contains the Heun-$D$ function which is the solution to the differential equation (\ref{eqR}) with $\epsilon=+1$. This equation in the limit of $a\rightarrow 0$ reduces to
\be
{\frac {{\rm d}^{2}}{{\rm d}{r}^{2}}}R \left( r \right) +{c^2}R \left( r
\right) +{\frac {{\frac {\rm d}{{\rm d}r}}R \left( r \right) }{r}}=0\label{Radialreduced},
\ee
for which the solutions are given by the Bessel functions  ${{\sl J}_{0}\left( {c}r\right)}$ and ${{\sl Y}_{0}\left( {c}r\right)}$. The Bessel function ${{\sl Y}_{0}\left( {c}r\right)}$ does not provide an oscillatory decaying behaviour similar to {that} of Heun-$D$ functions. {However,} the Bessel function ${{\sl J}_{0}\left( {c}r\right)}$  provides such a desired behaviour. 

So, we find an integral equation for the functions $f(c)$ and $g(c)$ which is
\be
\int_{0}^{\infty }J_0(cr)(f(c){{e}^{cx}}+g(c){{e}^{-cx}})\,{d}c=\frac{h_0}{\sqrt{r^2+x^2}}.
\ee

We find that the unique solutions to this integral equation for the functions $f(c)$ and $g(c)$ are given by
\be
f(c)=\frac{h_0}{2}\xi_1,\,\,g(c)=\frac{h_0}{2}\xi_2,
\ee
where $\xi_1$ and $\xi_2$ are constants and  $\xi_1+\xi_2=1$.
Furnished by all the necessary results, we have the most general solution for the convoluted metric function $H(r,x)$ which is given by
\be
H(r,x)=1+\frac{h_0}{2}\int_{0}^{\infty }{\cal H}_D\left(0,0,{a}^{2}c^2,0,\sqrt{{\frac {{a}^{2}+{r}^{2}}{{r}^{2}}}}\right)(\xi{{e}^{cx}}+(1-\xi){{e}^{-cx}})\,dc,\label{Hcfinal}
\ee
where $h_0$ and $\xi$ are two constants.

\section{The convoluted solutions in higher dimensional Einstein-Maxwell theory based on embedded Nutku instantons}
\label{sec:NutkuHigherD}

In this section, we find the general convoluted solutions for the embedding of Nutku geometry in higher than six dimensional Einstein-Maxwell theory. 
We consider the metric in $D$-dimensions as 
\be
ds^{2}=-H(r,x)^{-2}dt^{2}+H(r,x)^{\frac{2}{D-3}}(dx^2+x^2d\Omega_{D-6}^2+ds_{Nutku}^2),\label{N6DD}
\ee
where $d\Omega_{D-6}^2$ shows the metric on a $(D-6)$-dimensional unit sphere. We consider the gauge field with the non-zero component 
\be
A_t=\sqrt{\frac{D-2}{D-3}}\frac{1}{H(r,x)}.\label{gDD}
\ee
We find that all the $D$-dimensional Einstein and Maxwell equations are satisfied if the metric function $H(r,x)$ provides a solution to the partial differential equation
\be
{r} \sqrt{{ {{a}^{2}+{r}^{2}}}}\left({\frac {\partial ^{2}
}{\partial {x}^{2}}}H \left( r,x \right) + \frac{D-6}{x}\frac{\partial }{\partial x}H(r,x)\right)
+ (r^2+ {a}^{2})\left( {\frac {\partial ^{2
}}{\partial {r}^{2}}}H \left( r,x \right)  \right)
+ \left( {\frac {\partial }{\partial r}}H \left( r,x \right) 
 \right) r=0.\label{PDinD}
\ee
As in the six-dimensional case, we separate the coordinates in $H(r,x)$ by $H(r,x)=1+R(r)X(x)$. The partial differential equation (\ref{PDinD}) then separates into two ordinary differential equations for $R(r)$ and $X(x)$. The differential equation for the radial function $R(r)$ is given by (\ref{eqR}) where the solutions are given in terms of Heun-$D$ functions and the differential equation for $X(x)$ is
\be
{\frac {{d}^{2}}{{d}{x}^{2}}}X \left( x \right) -\epsilon c^2 {X}
\left( x \right)+{\frac{D-6}{x}}{\frac{d{X(x)}}{d{x}}} =0.\label{XeqD}
\ee
The solutions to (\ref{XeqD}) with $\epsilon=1$ for $D>6$ are given by
\be
X(x)=x^{\frac{7-D}{2}}I_{\frac{D-7}{2}}(cx)+x^{\frac{7-D}{2}}K_{\frac{D-7}{2}}(cx),
\ee
in terms of modified Bessel functions. As a result, we can write the most convoluted solution for the metric function in $D$ dimensions $H_D(r,x)$ as
\be
H_D(r,x)=1+\int_0^\infty {\cal H}_D\left(0,0,{a}^{2}c^2,0,\sqrt{{\frac {{a}^{2}+{r}^{2}}{{r}^{2}}}}\right)(f_D(c)I_{\frac{D-7}{2}}(cx)+g_D(c)K_{\frac{D-7}{2}}(cx))\,x^{\frac{7-D}{2}}{d}c,\label{HcD}
\ee
where $f_D(c)$ and $g_D(c)$ are two arbitrary functions in terms of separation constant $c$. We need to fix these two arbitrary functions to find a closed form for the metric function $H(r,x)$ in $D$-dimensions. As in the case of six-dimensional theory, we consider the limit of $a\rightarrow 0$, where the Nutku space reduces to $D^2 \otimes T^2$, with the line element (\ref{Na0}). In this limit, we find an exact analytical solutions to $D$-dimensional Einstein-Maxwell theory with the line element
\be
d{\cal S}^{2}=-H_{D0}(r,x)^{-2}dt^{2}+H_{D0}(r,x)^{\frac{2}{D-3}}(dx^2+x^2d\Omega_{D-6}^2ds_{a=0}^2),\label{NDDa0}
\ee
and the Maxwell gauge field 
\be
{\cal A}_t(r,x)=\sqrt{\frac{D-2}{D-3}}\frac{1}{H_{D0}(r,x)},\label{gDa0}
\ee
where the exact analytic metric function $H_{D0}(r,x)$ is
\be
H_{D0}(r,x)=1+\frac{h_{D0}}{{(r^2+x^2)^\frac{D-5}{2}}}\label{HD0},
\ee
where $h_{D0}$ is a constant.  To fix the functions $f_D(c)$ and $g_D(c)$, we demand that the metric (\ref{N6DD}) and the gauge field (\ref{gDD}), must reduce to the exact analytical metric (\ref{NDDa0}) and the gauge field (\ref{gDa0}), respectively, in the limit of $a \rightarrow 0$.  In other words, these requirements imply that the 
convoluted metric function $H_D(r,x)$ in equation (\ref{HcD}), must be equal to the exact analytic metric function $H_{D0}(r,x)$ in equation (\ref{HD0}), in the limit of $a \rightarrow 0$. As in the case of six-dimensional solutions, the integrand of the convoluted metric function $H_D(r,x)$ in equation (\ref{HcD}) contains the Heun-$D$ function which is the solution to the differential equation (\ref{eqR}) with $\epsilon=+1$. This equation in the limit of $a\rightarrow 0$ reduces to equation (\ref{Radialreduced}) 
for which the solutions are given by the Bessel functions  ${{\sl J}_{0}\left( {c}r\right)}$ and ${{\sl Y}_{0}\left( {c}r\right)}$.  

As a result, we find an integral equation for the functions $f_D(c)$ and $g_D(c)$ which is
\be
\int_{0}^{\infty }J_0(cr)(f_D(c)I_{\frac{D-7}{2}}(cx)+g_D(c)I_{\frac{D-7}{2}}(cx))x^{\frac{7-D}{2}}\,{d}c=\frac{h_{D0}}{{(r^2+x^2)}^{\frac{D-5}{2}}}.
\ee
After lengthy calculations, we find that the unique solutions to this integral equation for the functions $f_D(c)$ and $g_D(c)$ are given by
\be
f_D(c)=0\,\, ,g_D(c)=\alpha_D{h_{D0}}c^{\frac{D-5}{2}},
\ee
where $\alpha_D$'s are given by $\alpha_7=1,\,\alpha_8=\sqrt{\frac{2}{\pi}},\,\alpha_9=\frac{1}{2},\,\alpha_{10}=\frac{1}{3}\sqrt{\frac{2}{\pi}},\,\cdots$.
So, we can write the most general solution for the convoluted metric function $H_D(r,x)$ which is given by
\be
H_D(r,x)=1+\alpha_D{h_{D0}}\int_{0}^{\infty }{\cal H}_D\left(0,0,{a}^{2}c^2,0,\sqrt{{\frac {{a}^{2}+{r}^{2}}{{r}^{2}}}}\right)K_{\frac{D-7}{2}}(cx)\,x^{\frac{7-D}{2}}\,c^{\frac{D-5}{2}}\,dc.\label{HDcfinal}
\ee
We also note that asymptotically, a two-dimensional submanifold of our solutions (\ref{N6D}) and (\ref{N6DD}) represents the handle of {an} Ellis wormhole \cite{Ell73}.

\section{The cosmological convoluted solutions to the Einstein-Maxwell theory with a cosmological constant}
\label{sec:NutkuCosmo}
We consider the Einstein-Maxwell theory with a cosmological constant in six dimensions, where the metric function $H$ also depends explicitly on time coordinate 
\be
ds^{2}=-H(t,r,x)^{-2}dt^{2}+H(t,r,x)^{2/3}(dx^2+ds_{Nutku}^2).\label{N6Dcosmo}
\ee
We also consider the Maxwell gauge field as
\be
A_t(t,r,x)=\frac{2\sqrt{3}}{3H(t,r,x)}.\label{gcosmo}
\ee
The Einstein equations and Maxwell equations in presence of cosmological constant lead to a second order partial differential equation for $H(t,r,x)$, which is given by
\be
 \,{r} \left( \left( \frac{5}{3} \left( {\frac {\partial H}{
\partial t}}  \right) ^{2}-{\frac {3\,{
\Lambda}}{2}} \right)   H   ^{5/3}+
\, \left( {\frac {\partial ^{2}H}{\partial {t}^{2}}}  \right)   H  ^{8/3}-\,
{\frac {\partial ^{2}H}{\partial {x}^{2}}}
 \right) \sqrt {{ {{a}^{2}+{r}^{2}}  }}-\left( {a}^{2}+{r
}^{2} \right) {\frac {\partial ^{2}H}{\partial {r}^{2}}} -r{\frac {\partial H }{\partial r}}
  =0. 
\label{cosmoeq}
\ee
The form of partial differential equation (\ref{cosmoeq}) leads us to separate the coordinates as 
\be H(t,r,x)=1+T(t)+R(r) X(x). \label{sep}\ee 
We find three ordinary uncoupled differential equations for $R(r)$, $X(x)$ and $T(t)$ functions. The partial differential equations for $R(r)$ and $X(x)$ are given by (\ref{eqR}) and (\ref{eqX}) and the solutions to the differential equation for $T(t)$ are
\be
T(t)=\alpha t+\beta,
\ee 
where $\alpha=3\sqrt{\frac{\Lambda}{10}}$ and $\beta$ is an arbitrary constant. Requiring in the limit of $\Lambda \rightarrow 0$, the solution (\ref{sep}) for the metric function $H(t,r,x)$ approaches to the metric function $H(r,x)$ in asymptotically flat spacetime, yields $\beta=0$. Moreover, we find the following exact analytical non-stationary solutions to Einstein-Maxell theory with {a cosmological constant}
\be
d{\cal S}^{2}=-H_0(t,r,x)^{-2}dt^{2}+H_0(t,r,x)^{2/3}(dx^2+ds_{a=0}^2),\label{N6Da0cosmo}
\ee
and the time-dependent Maxwell gauge field 
\be
{\cal A}_t(t,r,x)=\frac{2\sqrt{3}}{3H_0(t,r,x)},\label{ga0cosmo}
\ee
where the exact analytic metric function $H_0(t,r,x)$ is given by
\be
H_0(t,r,x)=1+3\sqrt{\frac{\Lambda}{10}}t+\frac{h_0}{\sqrt{r^2+x^2}}\label{H0cosmo},
\ee
and $h_0$ is a constant. We note that $ds_{a=0}^2$ in (\ref{N6Da0cosmo}) is given by (\ref{Na0}).
We then can find the most general convoluted cosmological non-stationary solutions in six-dimensional Einstein-Maxwell theory with a cosmological constant where the metric function $H(t,r,x)$ approaches to $H_0(t,r,x)$ in the limit of $a\rightarrow 0$. The metric function $H(t,r,x)$ is equal to 
\be
H(t,r,x)=1+3\sqrt{\frac{\Lambda}{10}}t+\frac{h_0}{2}\int_{0}^{\infty }H_D\left(0,0,{a}^{2}c^2,0,\sqrt{{\frac {{a}^{2}+{r}^{2}}{{r}^{2}}}}\right)(\xi{{e}^{cx}}+(1-\xi){{e}^{-cx}})\,dc,\label{Hcfinalcosmo}
\ee
where $h_0$ and $\xi$ are two constants. 
The result (\ref{Hcfinalcosmo}) for the metric function $H(t,r,x)$ in six-dimensions can simply be generalized to solutions in $D$-dimensions, where we get
\be
H_D(t,r,x)=1+(D-3)\sqrt{\frac{2\Lambda}{(D-1)(D-2)}}t+\alpha_D{h_{D0}}\int_{0}^{\infty }{\cal H}_D\left(0,0,{a}^{2}c^2,0,\sqrt{{\frac {{a}^{2}+{r}^{2}}{{r}^{2}}}}\right)K_{\frac{D-7}{2}}(cx)\,x^{\frac{7-D}{2}}\,c^{\frac{D-5}{2}}\,dc.\label{HDcfinalcosmo}
\ee
 
As we notice, the general
solutions for the metric functions (\ref{Hcfinalcosmo}) and (\ref{HDcfinalcosmo}) describe the asymptotically dS spacetime. We can find the cosmological $c$-function for the solutions in the context of dS/CFT correspondence. As it is well known, for any $D$-dimensional asymptotically dS spacetime, one may define the c-function \cite{CC}
 \be
 c \sim (G_{\mu\nu}n^\mu n^\nu)^{1-D/2},
 \ee
 where $n^\mu$ is the unit vector along the time coordinate. If we have an expanding patch of dS, the flow of {the} renormalization group is {towards} the high energy region and the $c$-function is a monotonically increasing function in terms of {the} time coordinate. On the other hand, a contracting patch of dS, the flow of {the} renormalization group is {towards} the low energy region and the $c$-function is then a monotonically decreasing function in terms of {the} time coordinate. For example, the $c$-function for the $D$-dimensional solutions, given by the metric function (\ref{HDcfinalcosmo}), {describes} expanding patches of dS in different dimensions.

\bigskip

\section{Concluding remarks}

Inspired by {the} existence of helicoid-catenoid instantons in four {dimensions, we constructed} exact solutions to the five and higher dimensional Einstein-Maxwell theory with and without {a} cosmological constant. The solutions in five dimensions are given by the line element (\ref{metr5}), gauge field (\ref{gaugeD}) and the metric functions (\ref{sol5}).  We {discussed} the physical properties of the solution.  
{We also found} exact convoluted-like solutions {to six and} higher dimensional Einstein-Maxwell theory in which the metric functions are convoluted integrals of two special function with some measure functions.
We fix the measure functions for all the solutions by considering the solutions in some appropriate limits and comparing them with some other exact solutions in $D$-dimensions that are given by (\ref{N6D}) and (\ref{NDDa0}) with the metric functions (\ref{Hcfinal}) and  (\ref{HD0}), respectively. {We used} a special separation of variables to construct the solutions to Einstein-Maxwell theory with positive cosmological constant. In this case, the metric function depends on time and two spatial directions. The solutions are given by the metric (\ref{N6Dcosmo}) and gauge field (\ref{gcosmo}) where the cosmological metric functions in six and higher than six dimensions are given by (\ref{Hcfinalcosmo}) and (\ref{HDcfinalcosmo}), respectively.  We {constructed} the $c$-functions and notice that for all the cosmological convoluted solutions, the $c$-{function} is a monotonically increasing function in agreement with the $c$-theorem for asymptotically dS spacetimes. {As noted in \cite{Ali99}, the issue of singularities remains unresolved and needs further analysis.}

\bigskip

{\Large Acknowledgements}

This work was supported by the Natural Sciences and Engineering Research Council of Canada.

\end{document}